\documentclass[conference]{IEEEtran}  
\usepackage{booktabs} 
\usepackage{url}
\usepackage[keeplastbox]{flushend}

\usepackage{soul,color,graphicx,float}
\usepackage{eurosym,booktabs,multirow}
\usepackage{mathtools}
\usepackage{graphicx}
\usepackage{subfig}
\usepackage[euler]{textgreek}
\definecolor{Orange}{rgb}{1,0.5,0}

\usepackage{xspace}
\usepackage{ifthen}
\newboolean{authnotes}
\setboolean{authnotes}{true}

\ifthenelse{\boolean{authnotes}}
{
\newcommand{\usman}[1]{\footnote{{\bf Usman: #1}}}
\newcommand{\aftab}[1]{\footnote{{\bf Aftab: #1}}}
\newcommand{\tony}[1]{\footnote{{\bf Tony: #1}}}

\usepackage{draftwatermark}
\SetWatermarkText{}
}
{
\newcommand{\usman}[1]{}
\newcommand{\aftab}[1]{}
\newcommand{\tony}[1]{}
\usepackage{draftwatermark}
\SetWatermarkText{}

}

\newcommand{\fakeparagraph}[1]{\vspace{1mm}\noindent\textbf{#1}}

\newcommand{\lora}{LoRa\xspace}

\newcommand{\sigfox}{\textsc{SigFox}\xspace}

\makeatletter
\def\ps@IEEEtitlepagestyle{
  \def\@oddfoot{\mycopyrightnotice}
  \def\@evenfoot{}
}

\def\mycopyrightnotice{
  {\footnotesize
  \begin{minipage}{\textwidth}
  \centering
  \color{blue}
  Accepted to appear at IEEE GLOBECOM 2018.\\ Copyright~\copyright~2018  IEEE.  Personal  use  of  this  material  is  permitted.  Permission  from IEEE  must  be  obtained  for  all  other  uses,  in  any  current  or  future  media, including reprinting/republishing this material for advertising or promotional purposes, creating new collective works, for resale or redistribution to servers or lists, or reuse of any copyrighted component of this work in other work.
  \end{minipage}
  }
}

\title{How Agile is the Adaptive Data Rate Mechanism of LoRaWAN?}

\author{\IEEEauthorblockN{Shengyang Li\IEEEauthorrefmark{1}\IEEEauthorrefmark{2},
Usman Raza\IEEEauthorrefmark{1}, and
Aftab Khan\IEEEauthorrefmark{1}}\\ \vspace{0.25em}
\IEEEauthorblockA{\IEEEauthorrefmark{1}Toshiba Research Europe Ltd., Telecommunications Research Laboratory, Bristol, UK\\ \vspace{0.25em}
\IEEEauthorrefmark{2}University of Bristol, Bristol, UK\\
Email: sl14470.2014@my.bristol.ac.uk, usman.raza@toshiba-trel.com, aftab.khan@toshiba-trel.com}
}

\begin{document}
\maketitle

\begin{abstract}
The LoRaWAN based Low Power Wide Area networks aim to provide long-range connectivity to a large number of devices by exploiting  limited radio resources. The Adaptive Data Rate (ADR) mechanism controls the assignment of these resources to individual end-devices by a runtime adaptation of their communication parameters when the quality of links inevitably changes over time. This paper provides a detailed performance analysis of the ADR technique presented in the recently released LoRaWAN Specifications (v1.1). We show that the ADR technique lacks the agility to adapt to the changing link conditions, requiring a number of hours to days to converge to a reliable and energy-efficient communication state. As a vital step towards improving this situation, we then change different control knobs or parameters in the ADR technique to observe their effects on the convergence time. 
\end{abstract}

\section{Introduction}
The Low Power Wide Area (LPWA) networking technologies~\cite{razacomst} are recent outcomes of breakthroughs in communication technologies, as well as miniaturization and decreasing costs of the Internet of Things (IoT) devices. 
Their fast-paced adoption across all the seven continents aims to provide to tens of billions of devices a range in the order of tens of kilometers at a fraction of the cost and energy consumption of legacy technologies.
LoRaWAN  has established itself as one of the leading LPWA technologies in the past few years alongside \sigfox, Ingenu RPMA, and other cellular solutions such as NB-IoT and LTE-M. 
LoRaWAN is expected to connect a large number of static and mobile end-devices (EDs). 
Its support of the direct single-hop connection between EDs and the gateways obviates the need for any complex and expensive multi-hop mesh routing schemes. 

Nevertheless, LoRaWAN exploits a limited radio bandwidth available in the Sub-GHz part of the industrial, scientific and medical (ISM) band that is also shared among multiple co-existing technologies. 
In addition to this, the EDs and gateways must respect regional regulations related to the use of this spectrum, which restrict the time-on-air of transmissions. 
For these very reasons, efficient management of the scarce radio resources is essential for achieving very high \emph{network-wide} performance measured in terms of scale and goodput/reliability of a network. 
Apart from the problem of limited radio resources, the EDs are often deployed in very far-flung areas in challenging radio environments, resulting in very high variability in the link quality over time due to multiple different reasons such as obstructions, device mobility and environmental factors~\cite{cattani}. 
Thus, the responsibility of LoRaWAN stack extends well beyond sharing of the limited resources among a large number of EDs. 
It must also include intelligent mechanisms capable of \emph{adapting} the communication settings of \emph{individual EDs} to cope up with the link changes and thereby deliver reliable connectivity at all times.

In this paper, we study an \emph{Adaptive Data Rate (ADR)} mechanism that is responsible for the radio resource management and runtime link adaptation for individual EDs in LoRaWAN. 
This mechanism is recently updated in the LoRaWAN specifications v1.1~\cite{lorawanv11} and is gradually making its way into commercial deployments. 
The performance of this mechanism is yet to be fully understood. 
Therefore, we make a timely contribution in this paper by evaluating the performance of the ADR mechanism as a first step towards exposing different limitations and improving the underlying technique. 
Unlike most other studies~\cite{lancasterlorascale, khaled, flora,explora, 8114467} that focus on \emph{network-wide} performance metrics related to scalability, reliability, fairness, and throughput, we focus on performance that the individual EDs receive while running the ADR mechanism. 
We provide an in-depth study of the agility of the ADR mechanism in adapting the communication settings of EDs in response to link changes. 

In this paper, we also provide very first insights into the runtime performance of the official ADR algorithm under dynamic link conditions and various network sizes. 
Our detailed results reveal that if link conditions change or network size becomes too large, the convergence time of ADR mechanism to a communication setting that provides good reliability and low energy consumption is quite high. 
A large number of packets are therefore lost, motivating improvements in its design. 
We then investigate if changing different control knobs, specifically ADR initialization parameters, can improve the convergence rate, and if so to what extent. 
We also provide a brief discussion of the useful insights gained through this study that can help improve future ADR algorithms.

\begin{figure*}[t]%
    \centering
    \subfloat[ADR mechanism on ED side]{{\includegraphics[width=1\textwidth]{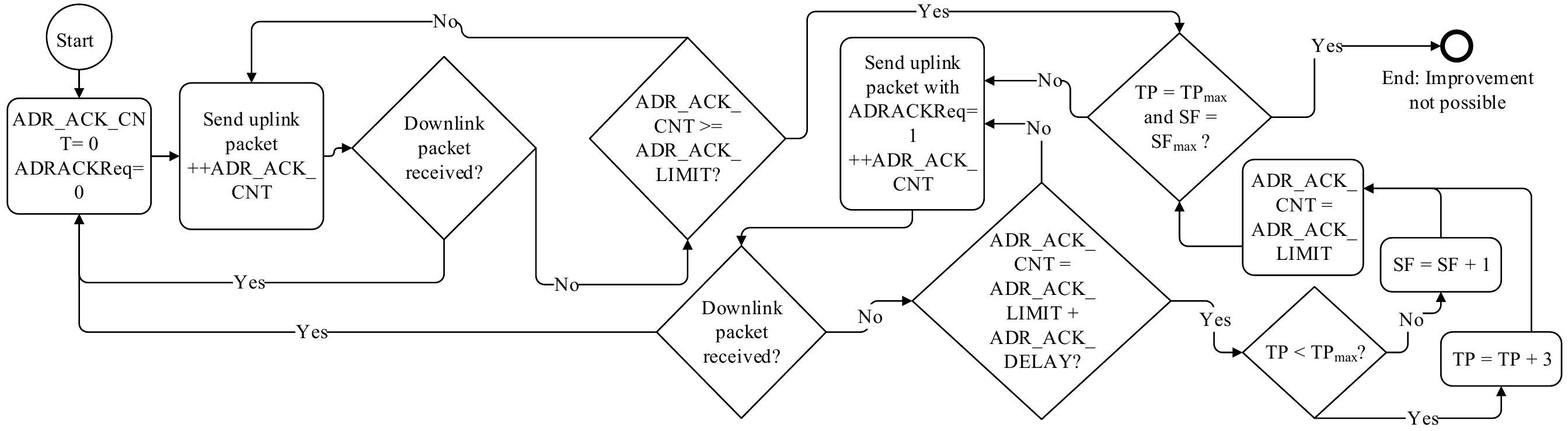} }}%
    \newline
    \subfloat[ADR mechanism on network side]{{\includegraphics[width=1\textwidth]{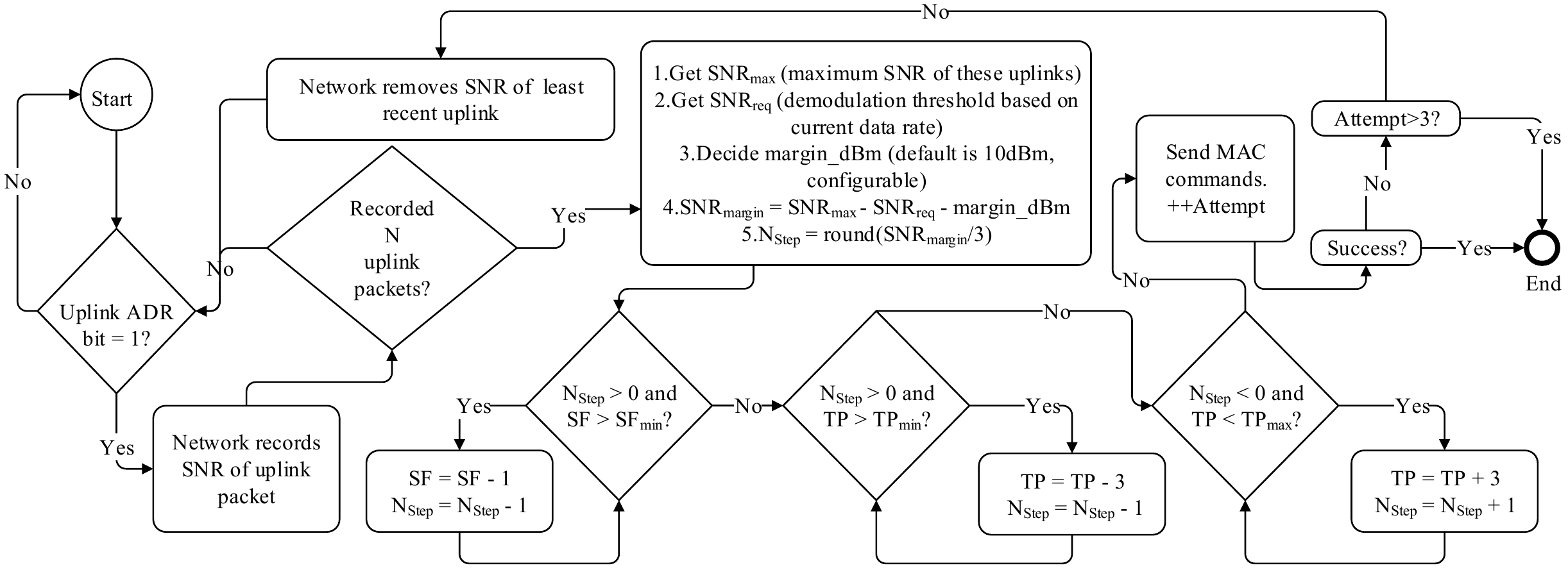} }}%
    \caption{ADR algorithm}%
    \label{fig:ADR}%
\end{figure*}

This paper is organized as follows: Section~\ref{sec:lorawan} presents a very brief primer on \lora and LoRaWAN immediately prior to the description of the ADR mechanism in Section~\ref{sec:adr}. 
Section~\ref{sec:simsetup} and Section~\ref{sec:eval} present simulation setup, the performance of the ADR mechanism and the impact of various factors on it. 
Section~\ref{sec:optimization} quantifies the extent to which the agility of the ADR mechanism can be improved by optimizing its different parameters. We then conclude the paper.   

\section{\lora and LoRaWAN}
\label{sec:lorawan}
\fakeparagraph{\lora}, a Chirp Spread Spectrum (CSS) technique, is the underlying PHY layer used by LoRaWAN, the upper network stack designed by LoRaWAN Alliance. 
The CSS technique supports multiple data rates. 
For a low data rate, it uses a large Spreading Factor (SF) that puts a high level of redundancy and amount of energy in a signal. 
Therefore, the signal can reach long distances and still retain enough strength to be successfully received. 
The same explanation holds for a signal sent at a high Transmission Power (TP). 
From the radio resource management point-of-view, the use of high SF values keeps radio medium busy for a long duration due to low data rates. 
Therefore, it is desirable to use the lowest possible SF that provides a good link between EDs and network. 
Different orthogonal SFs enable multiple successful receptions when used to send packets overlapping both in time and frequency channel. 
In the context of this paper, we restrict our discussion to SF and TP, the two control knobs adjusted by the ADR algorithm in pursuit of achieving long-distance, reliable, and energy-efficient communication. 

\fakeparagraph{LoRaWAN} defines the higher layer protocols and the network architecture that enable EDs to directly connect with the gateways using an ALOHA based multiple access scheme over the sub-GHz ISM bands. 
The gateways are then connected to the network servers that perform device authentication, downlink transmission scheduling, and execution of a part of ADR algorithm among many other important network-level functions. 
LoRaWAN mechanisms respect the regional regulation related to the use of the sub-GHz ISM spectrum, such as those governing maximum TP and duty cycles. 
This paper assumes operation of LoRaWAN in Europe where TP and duty cycle are limited to 14 dBm and 1\% respectively for the default frequency channels. 
The SF can be varied from 7 to 12 to adapt both the communication range and data rate. 
Three device classes are defined based on the application requirements for overall energy-efficiency and downlink communication latency. 
The most energy-efficient (and typically battery-powered) EDs are the Class A devices that experience the longest latency to receive downlink messages that are sent by the network only shortly after an uplink transmission.
The other device classes provide additional opportunities for receiving downlink messages at the expense of higher energy consumption. 

\section{Adaptive Data Rate Mechanism}
\label{sec:adr}
The ADR mechanism is part of LoRaWAN specifications. 
It aims to provide a fairly reliable and battery-friendly connectivity by adapting SF and TP to changes in link conditions. 
Both EDs and the network play an important role in this process. 

If an ED observes that a large number of consecutive uplink transmissions are not followed by a downlink response from the network, it assumes lost connectivity and resolves this issue by gradually stepping up its TP to the maximum before doing the same for SF. 
These measures gradually improve the robustness of the link. Figure~\ref{fig:ADR}a explains the full operation of EDs for adapting their TP and SF according to LoRaWAN Specifications v1.1. 
The two parameters namely ADR\_ACK\_LIMIT and ADR\_ACK\_DELAY control the number of uplink messages, after which if a downlink response is not received, an ED must increase either TP or SF. 
The value of these parameters along with the network size, deployment environment, and the amount of link fluctuations, all affect the time to converge to a state where ED is able to successfully re-establish a reliable link to the network. 
Section~\ref{sec:eval} provides a detailed analysis of these aspects.  

The EDs adapt communication setting to establish a reliable, but not necessarily an energy-efficient communication with the network. 
EDs can, however, request the network to step in and monitor the quality of uplink receptions from the recent past. 
If the link quality calculated over the last $N$ packets is too high compared to the minimum receiver sensitivity threshold, the network decides to reduce SF and/or TP. 
The new SF and TP values are set such that the expected signal-to-noise ratio of the future packets is above the minimum receiver sensitivity threshold by a pre-configured margin. 
Reduction in SF and TP would enable faster (high data rate) transmissions that consume less energy. 
Semtech, the organization that designed \lora, provides recommendations for implementing the network-side of the ADR algorithm, which is adopted by different operators as well as The Things Network, a popular crowd-sourcing LoRaWAN network. 
On the network-side, $N$, the minimum number of received packets that the network require to choose values of TP and SF, significantly affect the agility of the ADR algorithm as highlighted later in Section~\ref{sec:eval}.

\section{Simulation Setup}
\label{sec:simsetup}
We implement the ADR algorithm in LoRaWANSim~\cite{pop}, a discrete event simulator that already includes \lora as well as detailed LoRaWAN MAC protocol features including support for downlink traffic, retransmissions and support for Class A devices. 
Both EDs and the network are made capable of executing their sides of the algorithm shown in Figure~\ref{fig:ADR}, one of the contributions of this paper.  

To evaluate the ADR mechanism, we simulated networks consisting of a gateway with a varying number of randomly distributed EDs that transmit one packet every ten minutes on average. 
An urban scenario with the same path loss model as used by earlier works~\cite{pop,flora} is assumed. 
The EDs are tuned to use the default three central carrier frequencies in the g1 sub-band of the European sub-GHz ISM band, which is subject to a 1\% duty cycle limit. 
The simulation parameters used in following experiments are listed in Table \ref{Simulation P}.

Each experiment accounts for 12 days of simulated time and the reported results are averaged over 30 repetitions. 
The error bars in all the performance figures represent the standard deviations. 
\begin{table}[t!]
\caption {Simulation Parameters} 
\label{Simulation P} 
\begin{tabular}{lcc}
\toprule
Parameter               & {Value}                                    \\ 
\midrule
Communication range     & {670 m}                                    \\
Average message rate    & {10 minutes per message}                                    \\
Carrier frequency       & {g1 sub-band (868.1, 868.3, 868.5 MHz)}    \\
Bandwidth               & {125 kHz}                                  \\
Code rate               & {4/5}                                      \\
Spreading factor        & {7 to 12}                                  \\
Transmission power      & { \{2, 5, 8, 11, 14\} dBm }        \\
Path loss values \cite{lancasterlorascale}        & {d\textsubscript{0} = 40 m, \textgamma{} = 2.08, $\overline{L\textsubscript{pl}}$(d\textsubscript{0}) = 127.41 dB} \\
Channel variation level & {Low to High: \textsigma{} = \{0, 1.785, 3.57\} dB }          \\ 
\bottomrule
\end{tabular}
\end{table}


\section{Performance Evaluation}
\label{sec:eval}
We are mainly concerned with how quickly an ED, after experiencing a change in its link quality, converge to a state where it is assigned the right SF and TP values by the network. 
To quantitatively measure this, we define \emph{convergence time} as the duration from the change in the link quality until when the network receives enough number of packets required to compute the new  SF and TP values. 
We are also interested in the amount of energy consumed by the radio during this period.

Now, we evaluate the impact on the ADR mechanism due to various factors such as network scale, deployment environment, traffic type and link changes.

\subsection{Impact of Network Size}
Firstly, we start by looking at how network size affects the runtime performance of the ADR algorithm. 
For this particular experiment, we initialize networks of varying sizes and let the ADR algorithm run for all the EDs for some time so that the EDs acquire stable values of SF and TP. 
We then introduce to the networks additional 100 EDs and measure and report their convergence times.

\begin{figure}[t]%
    \centering
    \subfloat[Convergence time]{{\includegraphics[width=0.25\textwidth]{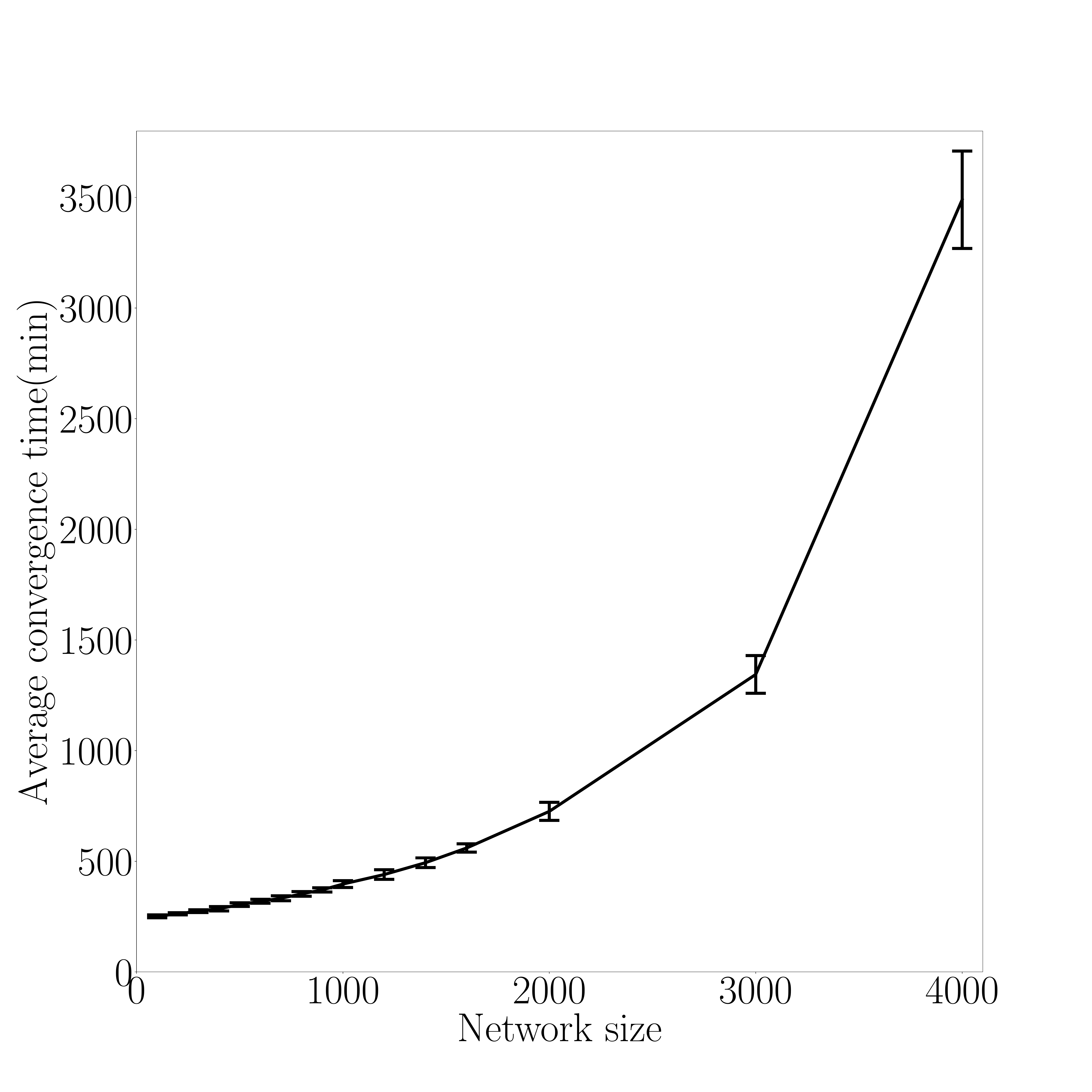} }}%
    \subfloat[Packet loss]{{\includegraphics[width=0.23\textwidth]{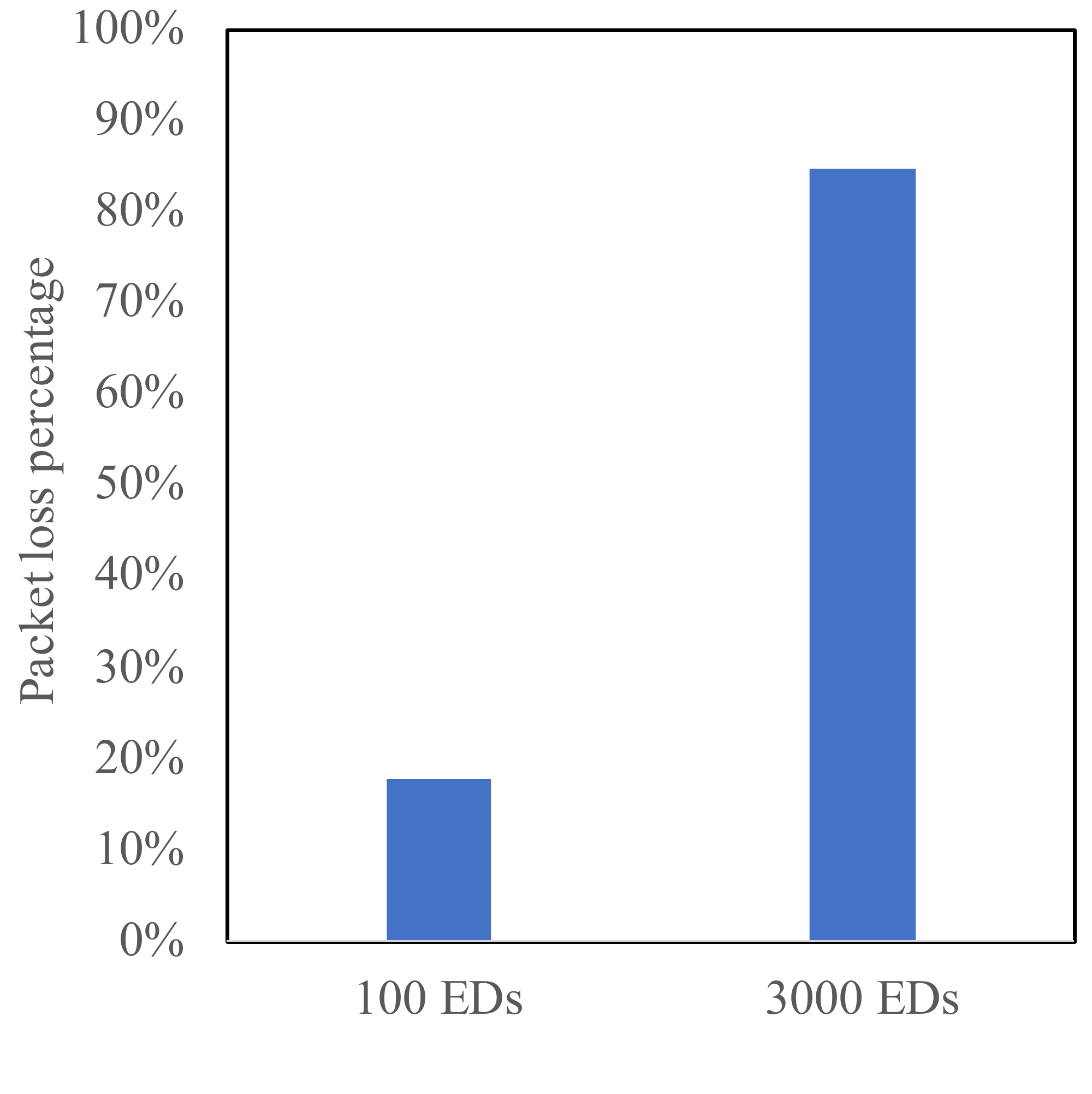} }}%
    \caption{Impact of network size on ADR algorithm.}%
    \label{networksize}%
\end{figure}

Figure~\ref{networksize}a shows that when the network size is increased, the convergence time increases as well from around 200 minutes for a  100-node network to more than 3000 minutes for a 4000-node network. 
The slow convergence in the large networks is due to a very high contention between a large number of uplink transmissions. 
This can be validated by Figure~\ref{networksize}b that shows that data loss due to collisions increases from approximately 17\% to 85\% as network scales from 100 to 3000 EDs. 
Thus, it takes more time for the network to receive $N$ transmissions from the EDs required to assign optimal TP and SF values. 
Furthermore, once the new values of TP and SF are calculated, the network is required to send downlink command messages to inform EDs. 
This is often not possible for a larger network due to the 1\% duty cycle limit on the transmissions from the gateway, resulting in additional delay.

\begin{figure}[t]
	  \vspace{-0.4cm}
      \centering
      \includegraphics[width=1.05\columnwidth]{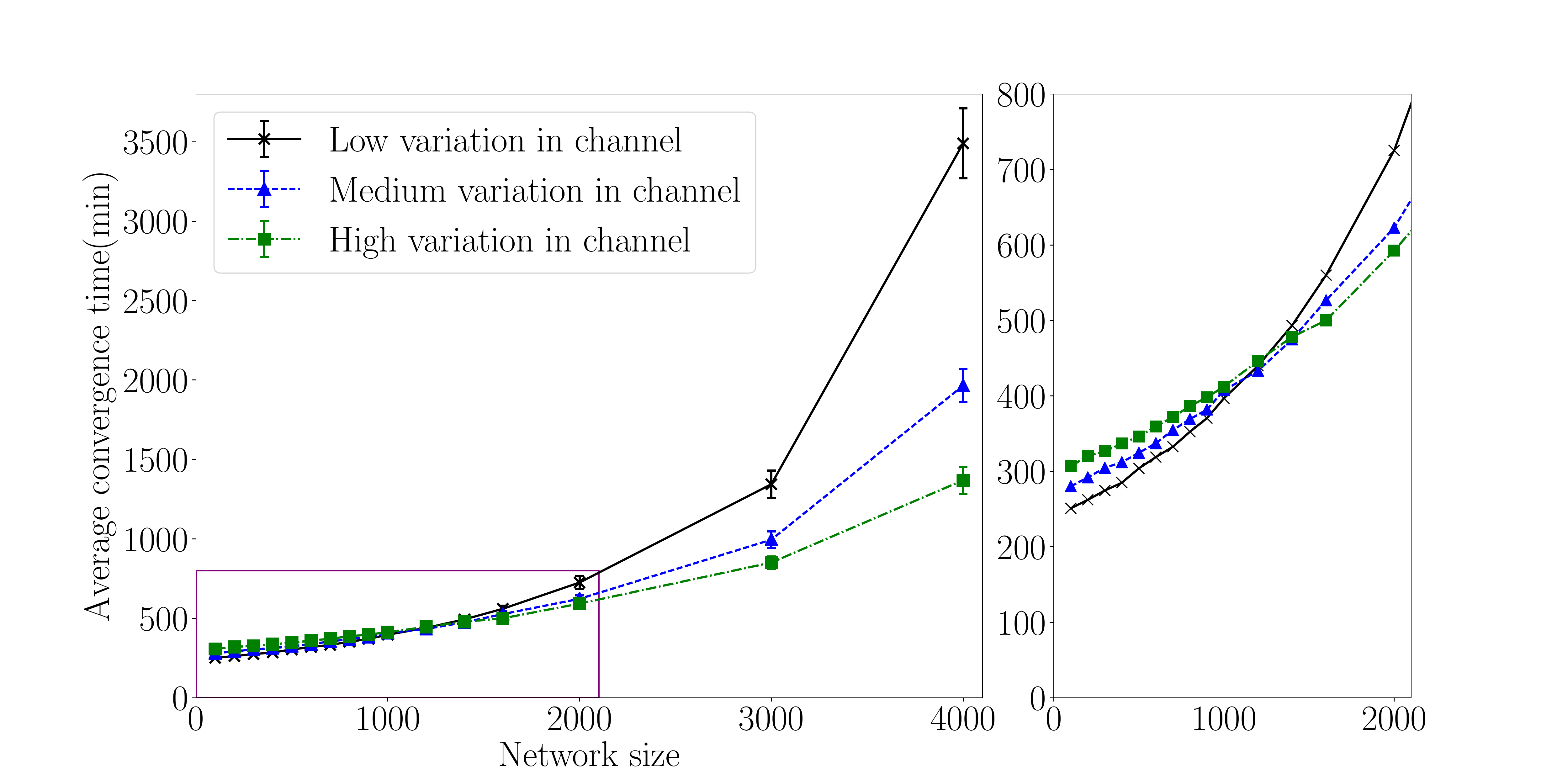}
      \vspace{0cm}
      \caption{Impact of channel condition on convergence time of ADR algorithm.}
      \label{channelcondition}
\end{figure}
\subsection{Impact of Deployment Environment}
Different deployment environments cause different levels of variation in the radio channel. 
We simulate three different scenarios related to low, medium and high levels of channel variation -- consistent with~\cite{flora}. 
This is achieved by changing the value of the standard deviation that accounts for the shadowing effect in the log-normal path loss model used in our simulations. 

Figure~\ref{channelcondition} shows the convergence time of the ADR algorithm under three different conditions. 
When network size is small, a high variation in channel slightly increases the convergence time of ADR algorithm. 

\begin{figure}[t]%
    \centering
    \subfloat[100 network size]{{\includegraphics[width=0.23\textwidth]{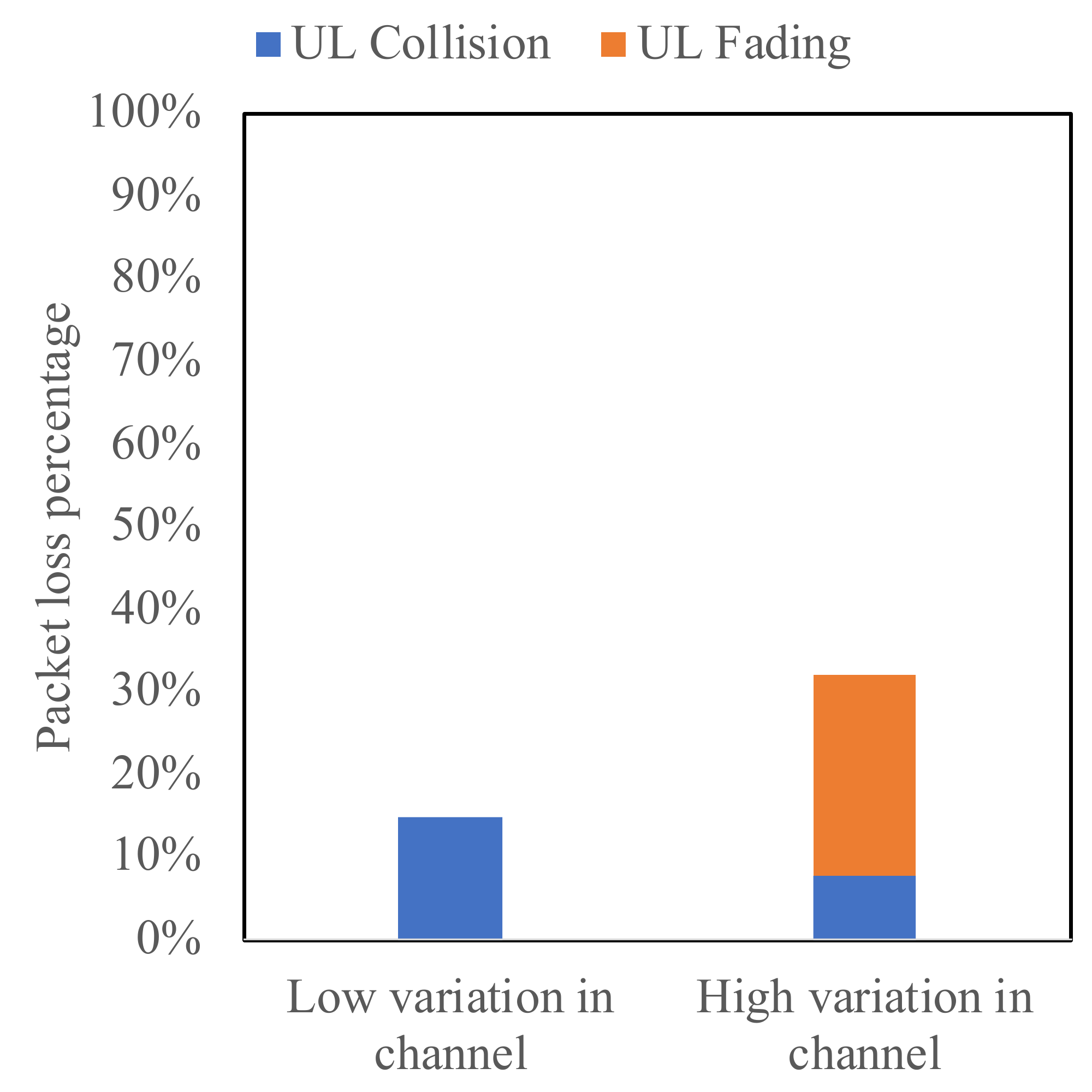} }}%
    \subfloat[3000 network size]{{\includegraphics[width=0.23\textwidth]{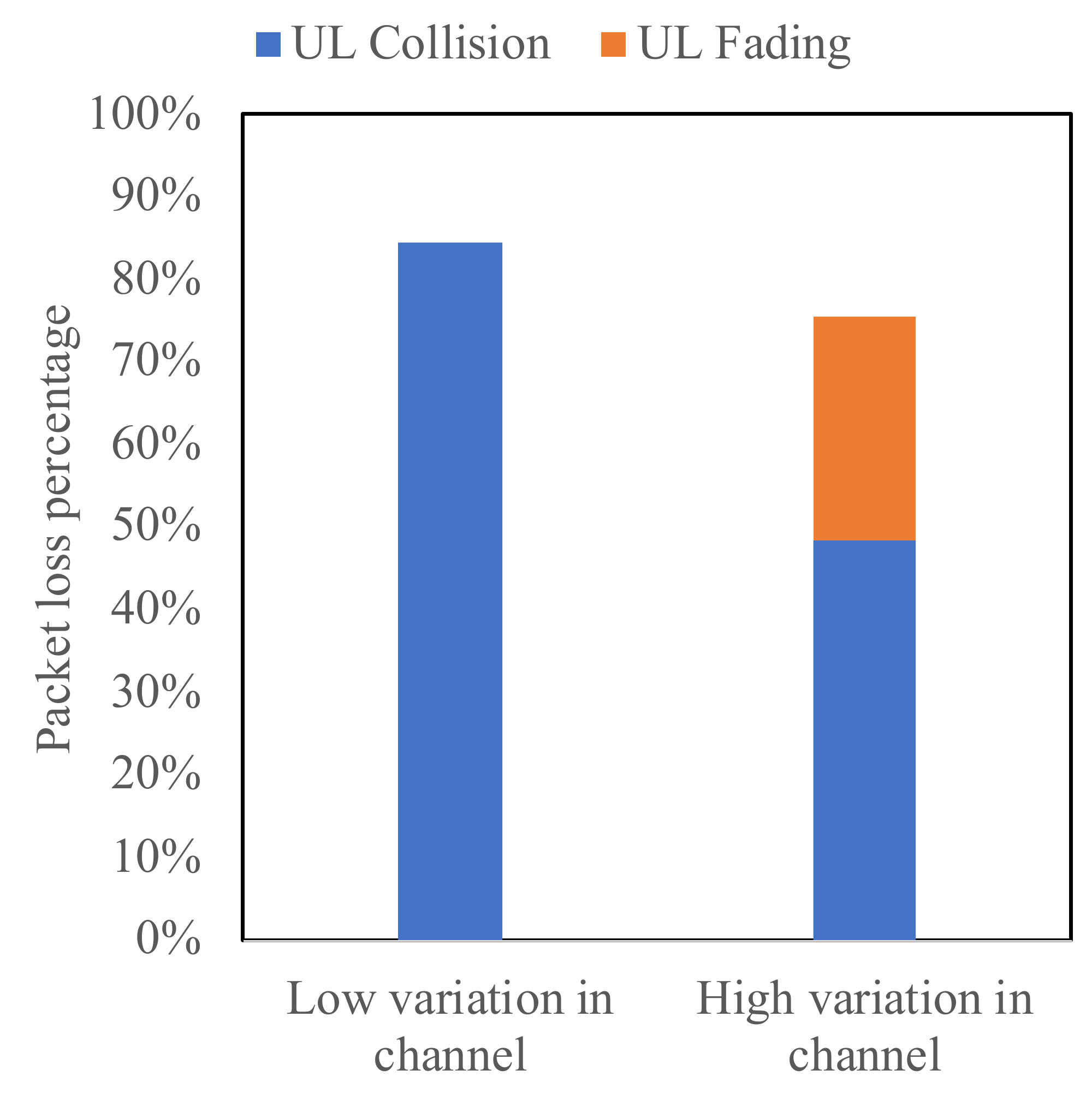} }}%
    \caption{Packet loss during ADR algorithm}%
    \label{packetloss}%
\end{figure}

Whilst one will naturally expect this to happen, higher variation in channel signifies that uplink packets are more likely suffering from losses due to fading. 
Figure~\ref{packetloss} highlights different reasons of packet loss. 
The total percentage packet loss under highly varying channel exceeds the total percentage packet loss under a less varying channel. 
However, collisions happen more often under low variation. 
High packet loss translates to longer convergence times.

\begin{figure}[t]%
\vspace{-0.88cm}
    \centering
    \subfloat[Increase in path loss]{{\includegraphics[width=0.25\textwidth]{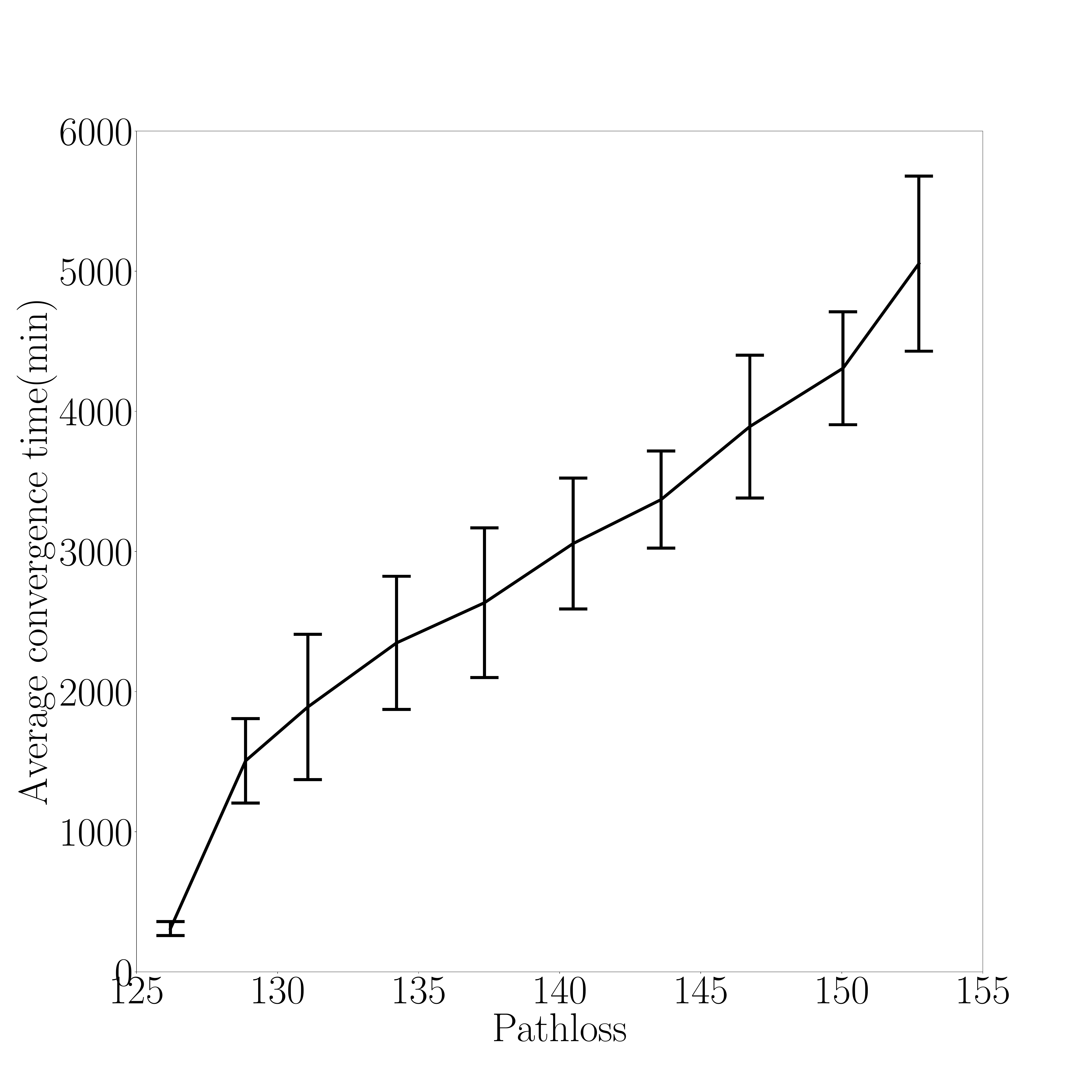} }}%
    \subfloat[Decrease in path loss]{{\includegraphics[width=0.25\textwidth]{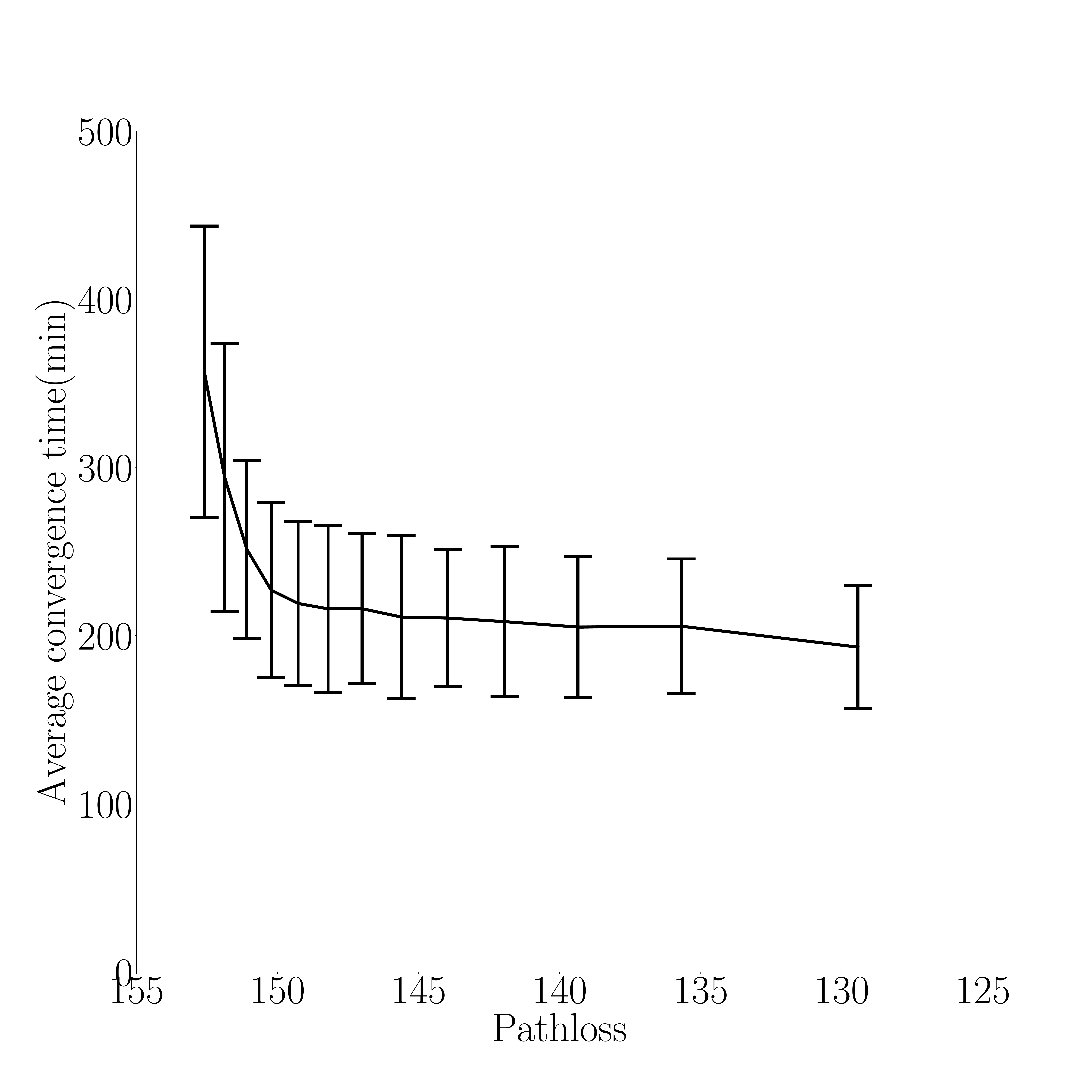} }}%
    \caption{Impact of change in link quality on convergence time for the ADR algorithm.}%
    \label{fig:linkquanlity}%
\end{figure}

When the network size is large, we can observe that higher variation in a channel does not necessarily lead to an increase in the convergence time. 
Rather, it actually reduces the convergence time. 
As shown in Figure~\ref{packetloss}b, the uplink collisions happen much more for low channel variation. 
Although uplink fading under high varying channel occurs more often than the one in the low varying channel as expected. 
But under the low varying channel, the increase in the uplink collisions is so high that the total packet loss outweighs the one under high varying channel.
One potential reason is that a high variation in channel introduces randomness in the received signal strength of the uplink packets.
This leads to a large difference between signal strengths of overlapping packets, resulting in successful decoding of the strongest signal due to the \textit{}{capture effect}. 

More uplink packets can therefore successfully reach the gateway. 
Another possible reason is high variation in channel eases the 
crowded network. 
High variations in channel cause some uplink packets to not reach the gateway, reducing the contentions for others that do.
Thus, the chances of collision reduce at gateway. 
Overall, the success of uplink packets reaching gateway is more influenced by the number of collisions rather than fading for large network size. 
This trend is reversed for network size.

\begin{figure}[t]%
    \centering
    \subfloat[Convergence time]{{\includegraphics[width=0.25\textwidth]{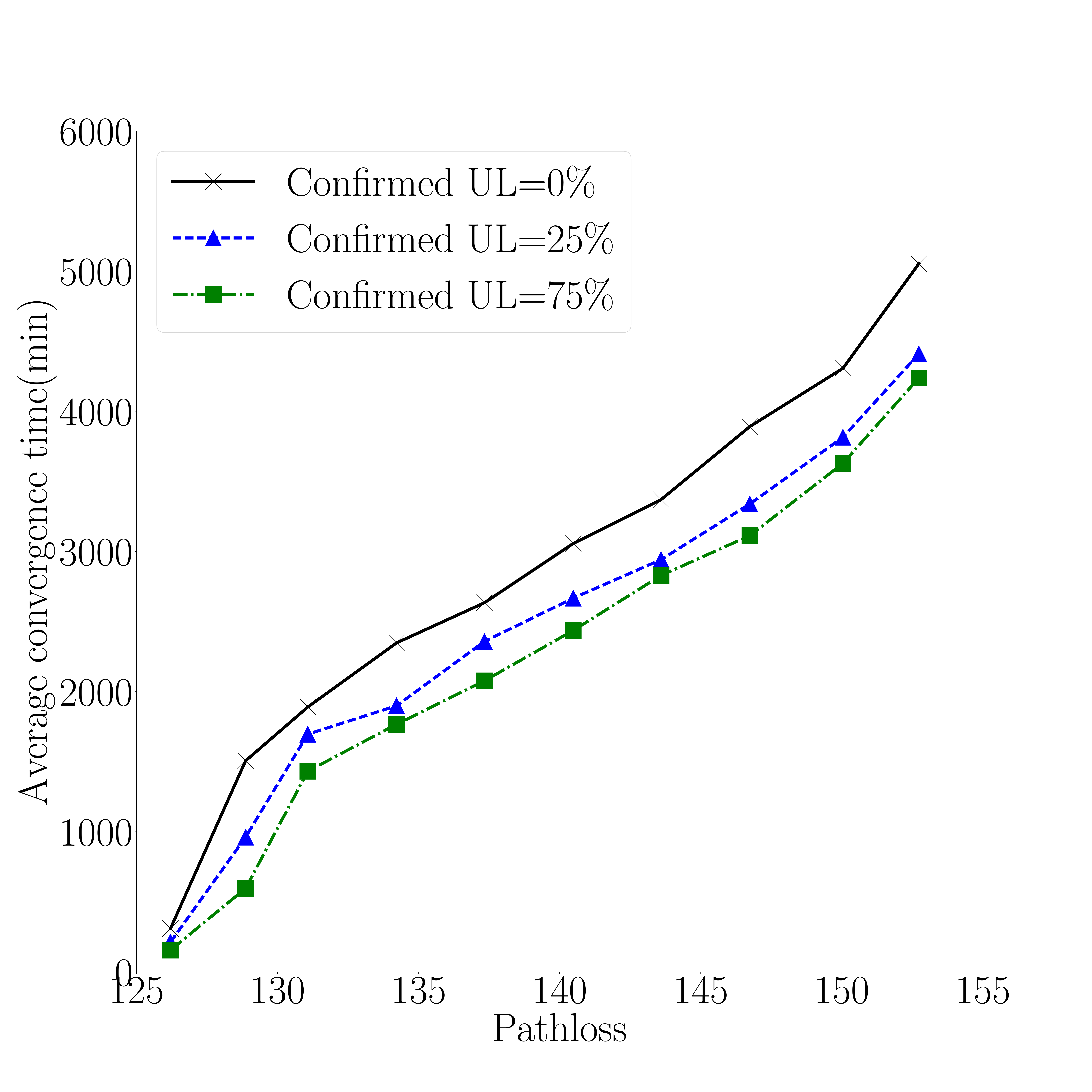} }}%
    \subfloat[Energy consumption]{{\includegraphics[width=0.25\textwidth]{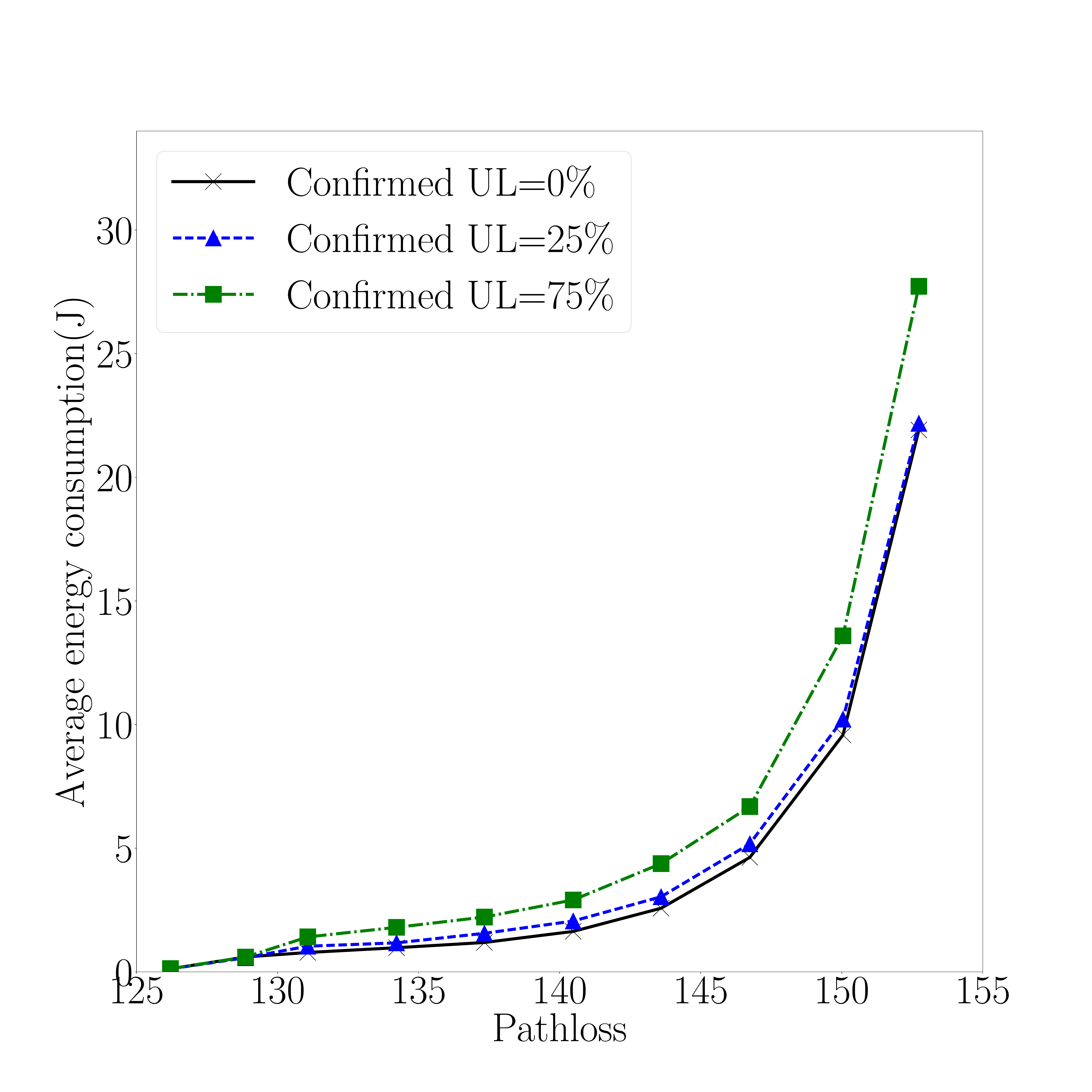} }}%
    \caption{Impact of confirmed uplinks on ADR algorithm}%
    \label{fig:ADR_uplink_cost}%
\end{figure}

\begin{figure}[t]%
    \centering
    \subfloat[Convergence time]{{\includegraphics[width=0.25\textwidth]{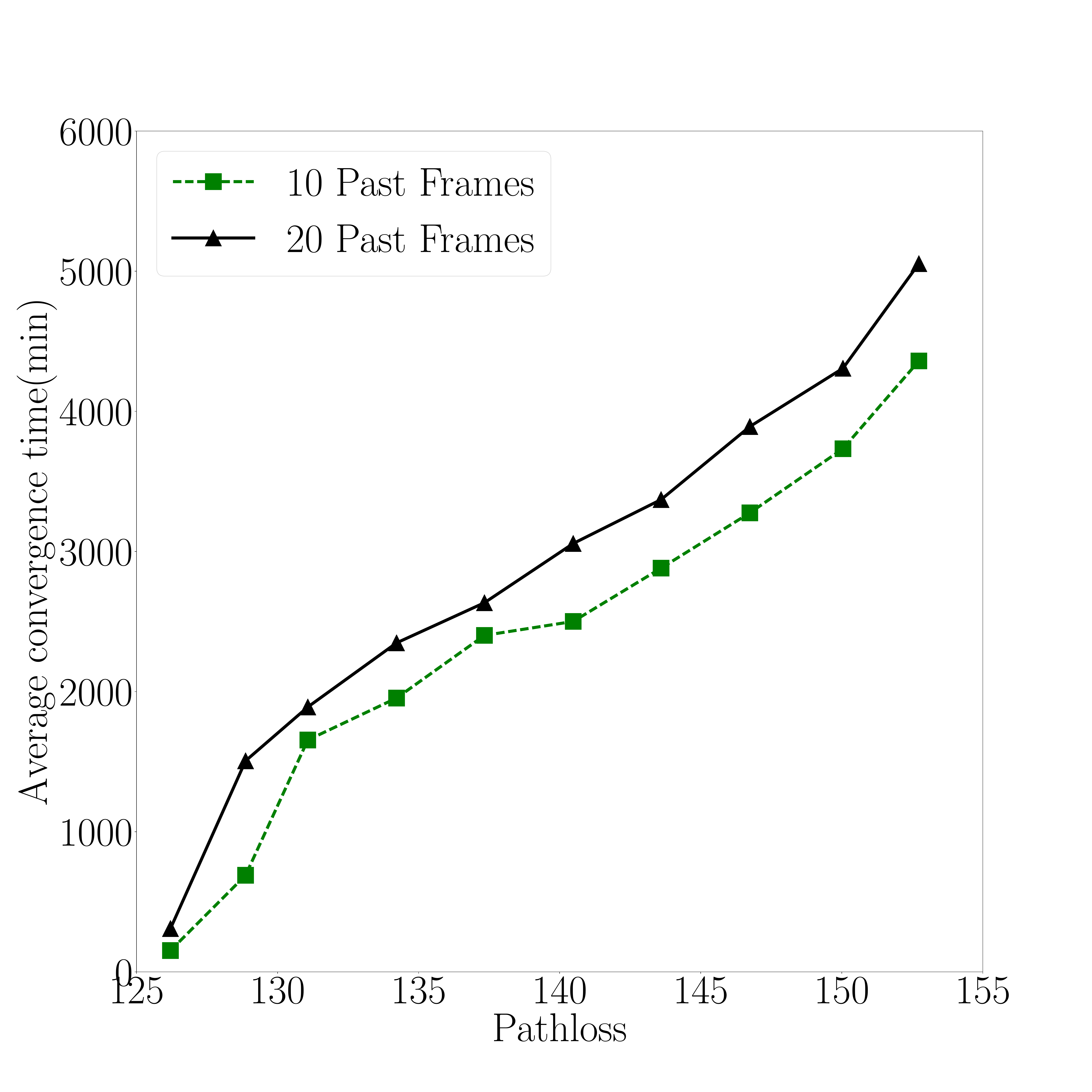} }}%
    \subfloat[Energy consumption]{{\includegraphics[width=0.25\textwidth]{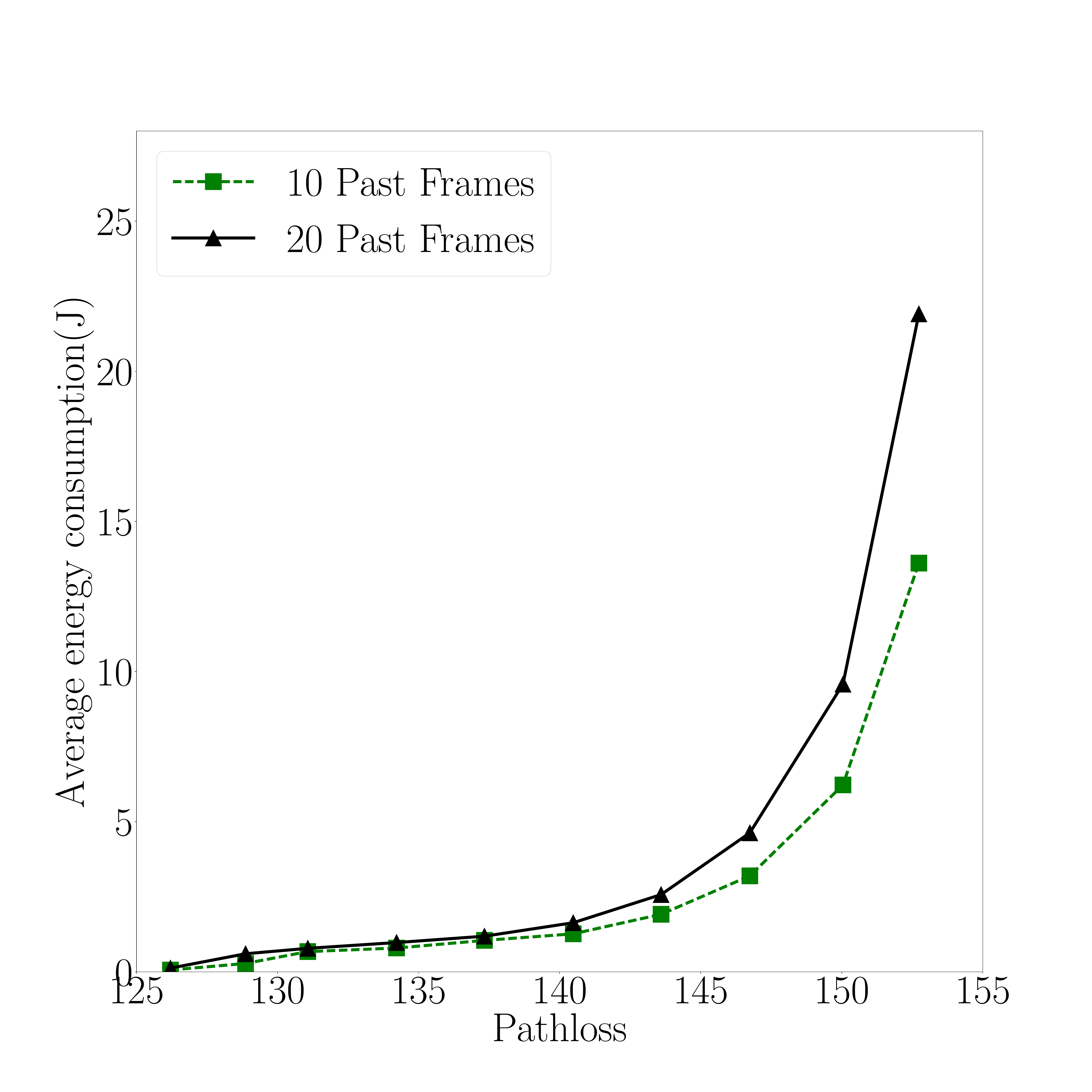} }}%
    \caption{Impact of number of past frames collected on ADR algorithm}%
    \label{fig:nrframes}%
\end{figure}

\subsection{Impact of Link Changes}

LoRaWAN is a technology of choice for various smart city applications that are often deployed in dense urban environments. 
Wireless links may often degrade or improve in such environments suddenly. 
Smart parking application is one example where the parking sensors often are obstructed by vehicles causing links to degrade. 
Now, we simulate such changes in communication environment by altering the mean path loss value of the communication between end devices and gateway. 
Figure~\ref{fig:linkquanlity}a simulates a case when the link quality of an ED degrades due to obstructions or mobility that increase the mean path loss. 
Figure~\ref{fig:linkquanlity}b, on the other end, shows a case where ED improves its link quality by reducing mean path loss.

Figure~\ref{fig:linkquanlity}a clearly shows when the link quality degrades, the time required by the ADR algorithm to converge to the right communication parameter setting increases significantly. 
This overhead mainly comes from the process running on ED to regain connectivity to the gateway. Unfortunately, this process requires EDs to lose sufficient number of sent packets before moving to higher SF or TP values.
In Figure~\ref{fig:linkquanlity}b, when node has a good link quality, initially ADR takes a slightly longer time, mainly because of channel variation and resulting additional packets due to loss of uplink packets. 
However as the link quality continues to improve, gateway continues to receive the packets for computing SF and TP. 
In our simulations, the convergence time approaches 200 minutes, the minimum time required to receive $N=20$ packets that are transmitted every 10 minutes on average.

\subsection{Impact of Traffic Type} 

Some application traffic in LoRaWAN can be of more value than the rest and therefore should be acknowledged by the network. 
For this purpose, LoRaWAN supports confirmed messages. 
We now are interested in how the convergence time is affected by the confirmed uplink traffic. 
Figure~\ref{fig:ADR_uplink_cost}a shows the effect of different percentage of uplink packets requiring ACK on the convergence time. 
If an ACK is lost, the frame will be retransmitted after a short time. 
Retransmissions improve link reliability, enabling the gateway to collect enough number of packets quickly, thus reaching optimal SF and TP earlier compared to the cases when all the messages are unconfirmed.

\begin{figure}[t]%
    \centering
    \subfloat[Convergence time]{{\includegraphics[width=0.25\textwidth]{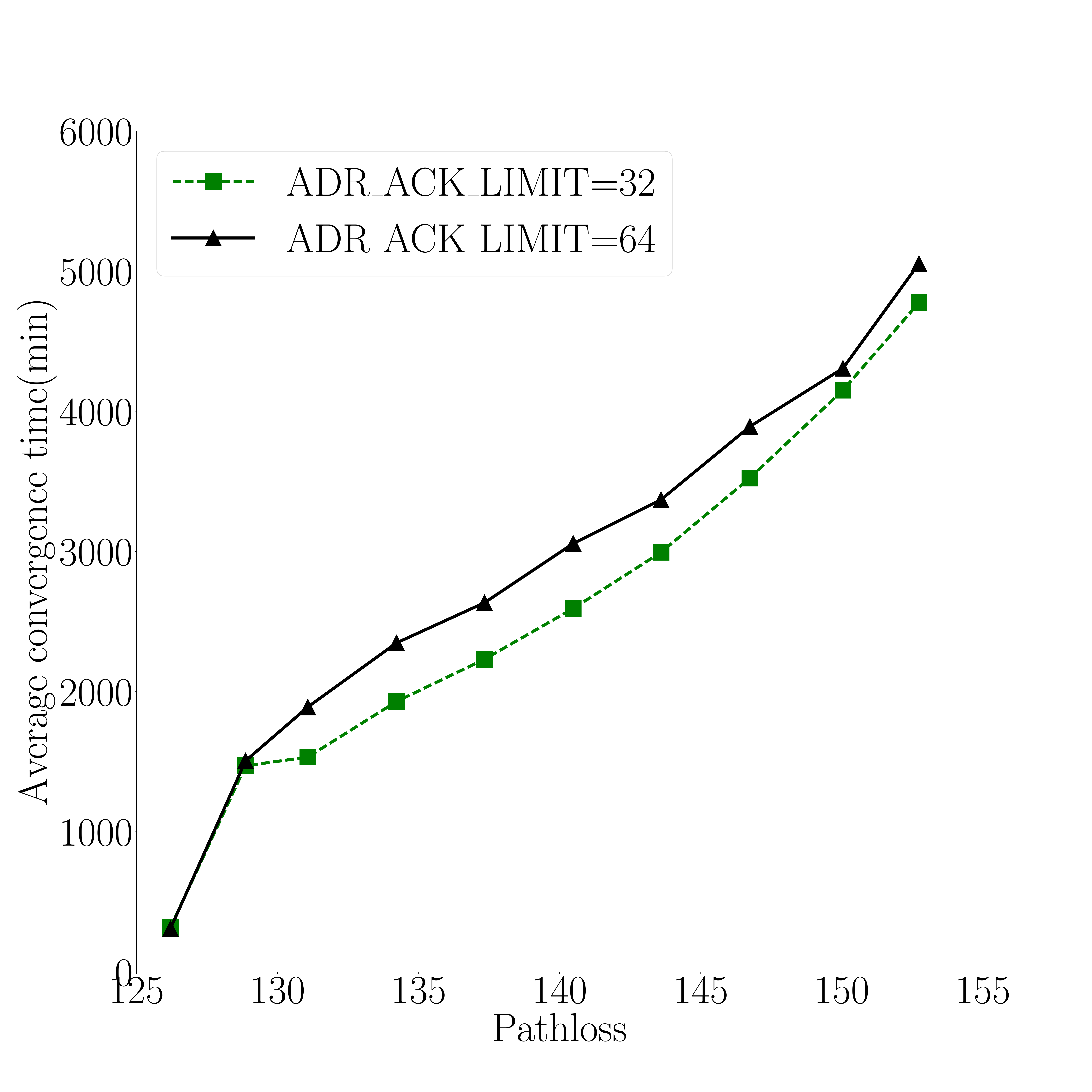} }}%
    \subfloat[Energy consumption]{{\includegraphics[width=0.25\textwidth]{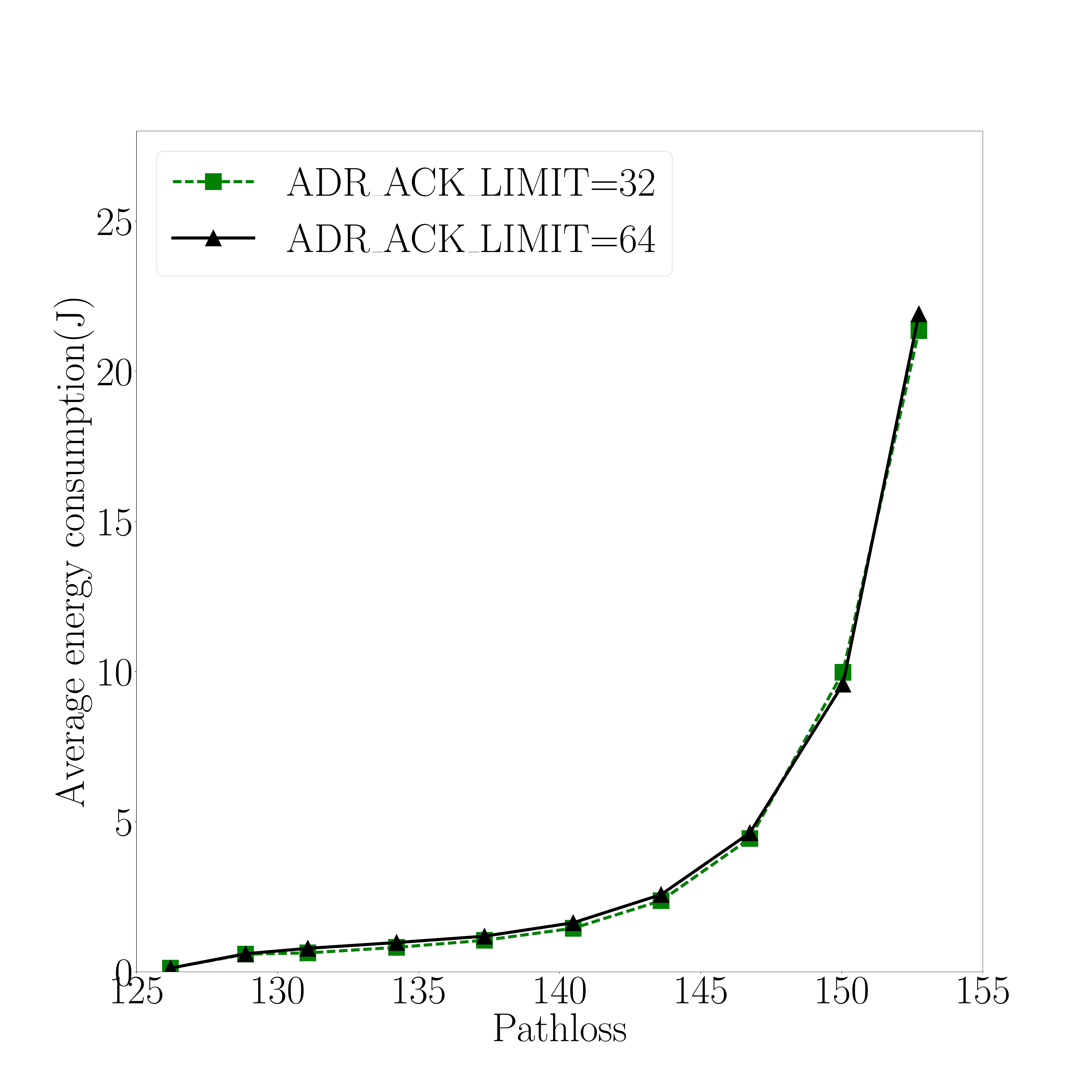} }}%
    \caption{Impact of ADR\_ACK\_LIMIT on ADR algorithm}%
    \label{fig:ADR_ACK_LIMIT}%
\end{figure}

\begin{figure}[t]%
    \centering
    \subfloat[Convergence time]{{\includegraphics[width=0.25\textwidth]{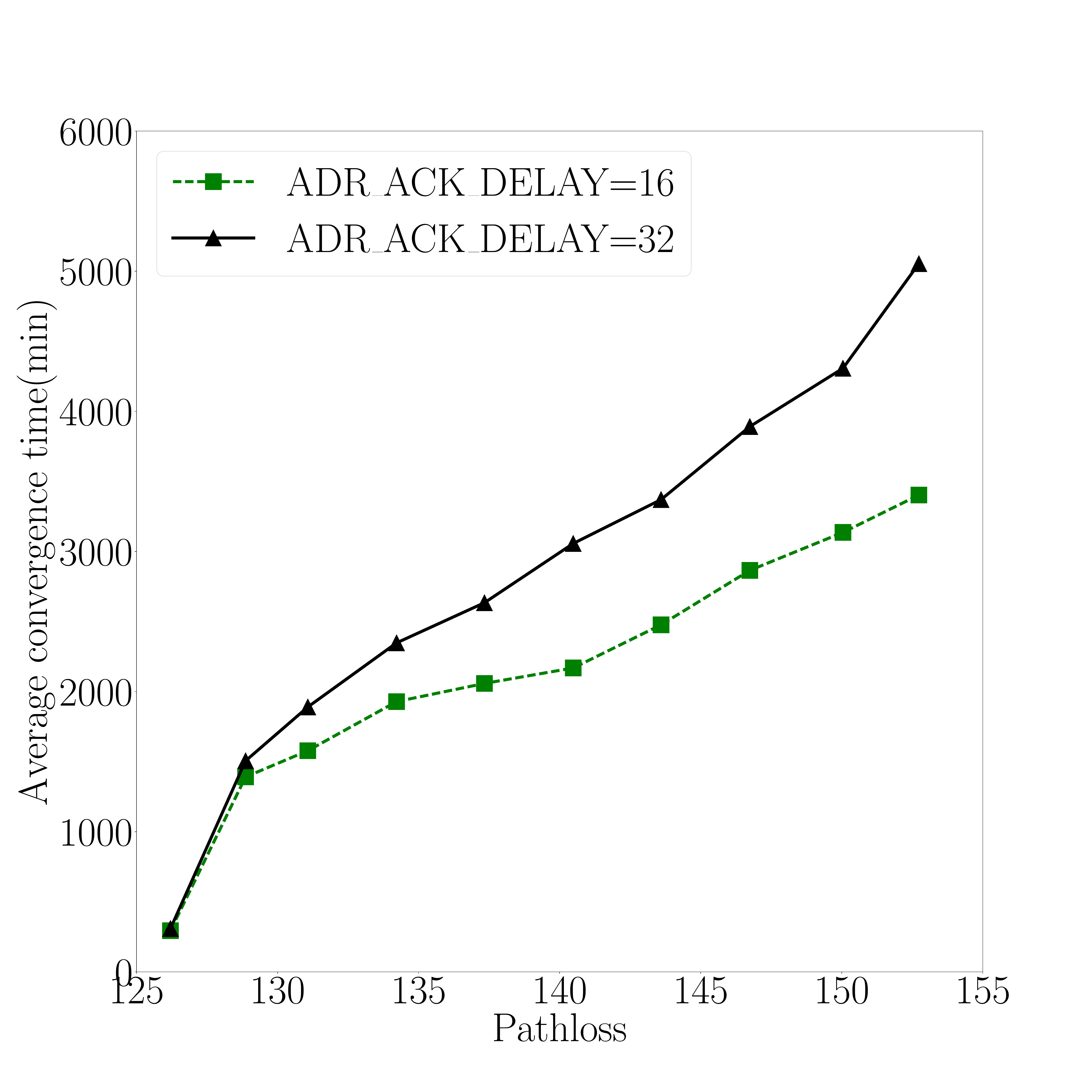} }}%
    \subfloat[Energy consumption]{{\includegraphics[width=0.25\textwidth]{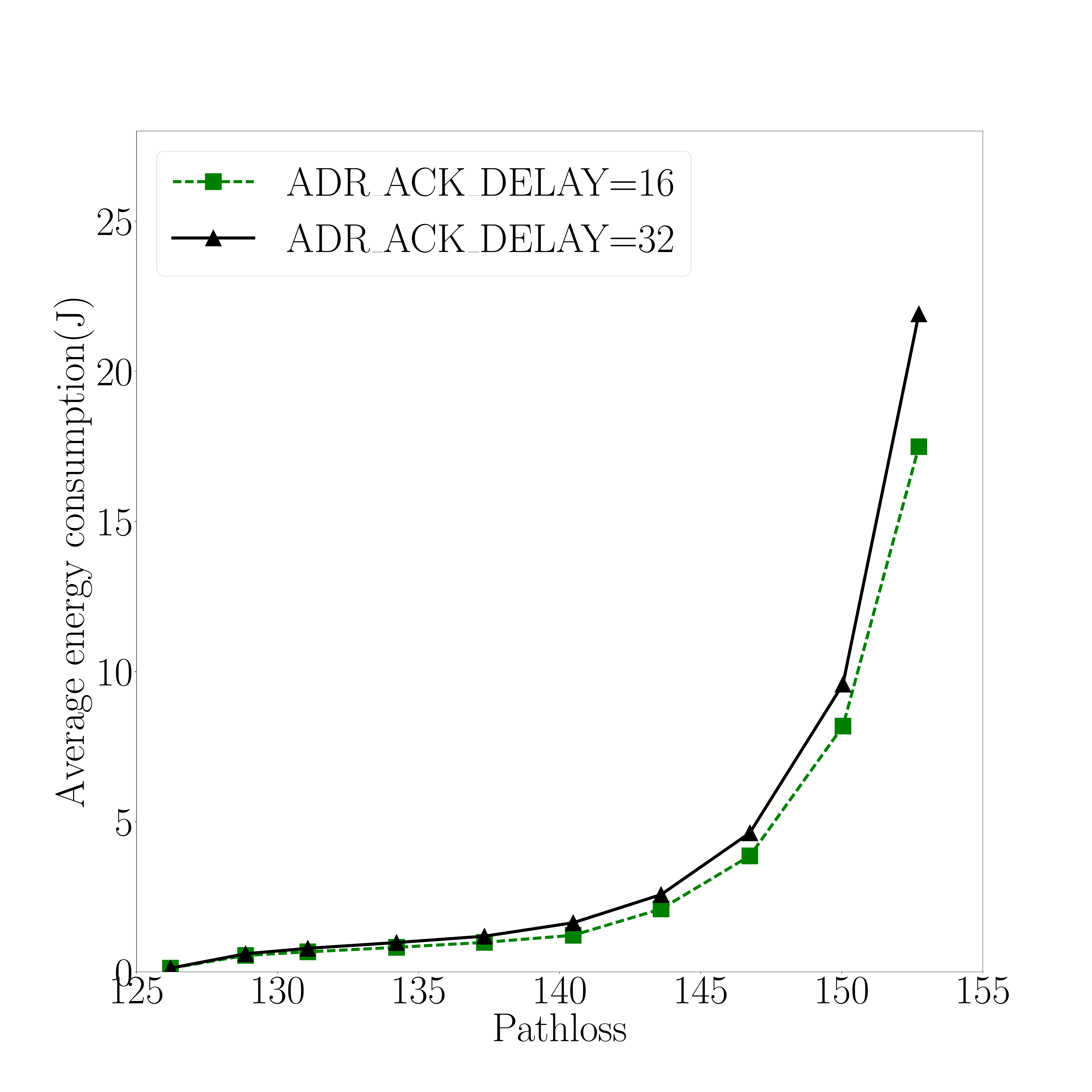} }}%
    \caption{Impact of ADR\_ACK\_DELAY on ADR algorithm}%
    \label{fig:ADR_ACK_DELAY}%
\end{figure}


\section{Optimization for the ADR Mechanism}
\label{sec:optimization}

\begin{table}[t!]
\caption {Tunable parameters in the ADR mechanism} 
\label{tp} 
\begin{tabular}{p{3cm}p{4.8cm}}
\toprule
Parameter               & {Description}                                    \\ 
\midrule
$N$     & {Number of packets required by the network to compute SF \& TP}                                    \\
ADR\_ACK\_LIMIT    & {Threshold on number of lost packets to force DL ACK}                                    \\
ADR\_ACK\_DELAY    & {Threshold on number of lost packets to increase SF or TP}    \\
\bottomrule
\end{tabular}
\end{table}

After learning that the convergence of ADR is slow, we now change different tunable parameters of the ADR algorithm (shown in Table~\ref{tp}) to analyze the corresponding effect on its performance. We consider $N$, the number of packets required by the network for ADR calculation, ADR\_ACK\_DELAY and ADR\_ACK\_LIMIT for this purpose.

From Figure \ref{fig:nrframes} to Figure \ref{fig:ADR_ACK_DELAY}, we show convergence time and energy consumption of ADR algorithm using different parameter settings. 
Both Figures \ref{fig:nrframes} and \ref{fig:ADR_ACK_LIMIT} show only marginal improvement in convergence time. 
This can mainly be attributed to the fact that most time is consumed by the node to regain connectivity. 
This is achieved by increasing transmission power and spreading factor step by step. 

Reducing the value of $N$ requires the gateway to collect less number of packets in order to decide the values of SF and TP. This, in principle, should reduce the convergence time. However, this does not bring significant reduction in the overall time. This is because if the link is really bad, nodes are still required to send and lose sufficient number of packets to increase their SF and TP gradually to a reliable communication setting. In this case, the overall convergence time is mainly dominated by regaining connectivity rather than collection of $N$ packets.

Alternatively, if we reduce value of ADR\_ACK\_LIMIT, this will speed-up the process of requesting downlink response (ADRACKReq). Nevertheless, if the link is still bad, the EDs may have to send multiple of ADR\_ACK\_DELAY packets to gradually step up to right settings of SF and TP. Therefore, tuning  $N$ and ADR\_ACK\_LIMIT will not result in a significant reduction in convergence time especially if the link quality degrades quite a lot. 

Figure \ref{fig:ADR_ACK_DELAY} shows a better improvement in both convergence time and energy consumption of the algorithm if we reduce ADR\_ACK\_DELAY. 
This is easy to understand if we look back at the functionality of ADR on the node side. Nodes will only increase either TP or SF every time when counter reaches (ADR\_ACK\_LIMIT + ADR\_ACK\_DELAY). 
Reducing ADR\_ACK\_DELAY means we decrease the duration of each individual step that increases either TP or SF. 
The resulting shorter duration will accumulate to speed-up the process of making link more robust against packet loses. Effectively, the shorter convergence time also brings the benefit of less energy consumption because of less number of transmission attempts.

\section{Discussion and Conclusion}
\label{sec:discussion}
The main objective of this paper was to understand the performance of the official ADR mechanism proposed by LoRaWAN Alliance in LoRaWAN specifications v1.1. We assessed the impact of different configurable parameters on the performance of the ADR mechanism that runs on both the EDs and the network. In this process, we attempted to answer several questions i.e., how different factors impact the convergence time.
We provide useful insights into improving the algorithm:

\begin{itemize}
\item The convergence of the ADR mechanism is slow, more significantly when the link quality degrades and EDs need to move from lower to higher value of SF or TP to regain connectivity. This suggests that ADR should promptly identify the onset of lost connectivity and then increase SF and TX power. 
\item The convergence time is more sensitive to ADR\_ACK\_DELAY compared to ADR\_ACK\_LIMIT. These observations should be taken into account while improving the ADR algorithm. 
\end{itemize} 

The slow convergence rate of the ADR mechanism also introduces higher energy consumption and packet losses. The lack of the necessary agility to adapt to changing link in our opinion is a good research challange for current LoRaWAN specifications. 
Our observations about the effect of different paramters on the convergence time provide a good direction towards proposing the next generation of ADR algorithms that, in addition to being adaptive by definition, must also be agile and more reliable.


\bibliographystyle{IEEEtran}
\bibliography{refs}
\end{document}